\documentclass[apj]{emulateapj}

\begin{document}

\title{Interpretation of a Variable Reflection Nebula Associated with HBC 340 and HBC 341 in NGC\,1333}

\author{S. E. Dahm\altaffilmark{1} \& L. A. Hillenbrand\altaffilmark{2}}

\altaffiltext{1}{U.S. Naval Observatory, Flagstaff Station, 10391 West Naval Observatory Road, Flagstaff, AZ 86005-8521, USA}
\altaffiltext{2}{Department of Astronomy, California Institute of Technology, Pasadena, CA 91125, USA}

\begin{abstract}
We present multi-epoch, $R-$band imaging obtained from the Palomar Transient Factory of 
a small, fan-shaped reflection nebula in NGC\,1333 that experiences prominent brightness
fluctuations. Photometry of HBC 340 (K7e) and HBC 341 (M5e), a visual pair of late-type, 
young stellar objects lying near the apex of the nebula, demonstrates that while both 
are variable, the former has brightened by more than two magnitudes following a deep 
local minimum in September 2014. Keck high dispersion (R$\sim$45,000-66,000), optical 
spectroscopy of HBC 340 suggests that the protostar is a spectroscopic binary 
(HBC 340Aa + HBC 340Ab). Both HBC 340 and HBC 341 exhibit strong H$\alpha$ and forbidden
line emission, consistent with accretion and outflow. We conclude that the brightness fluctuations 
in the reflection nebula represent light echos produced by varying incident radiation 
emanating from HBC 340. The short-term variability observed in the protostar is attributed 
to irregular accretion activity, while correlated, dipping behavior on a several hundred 
day time scale may be due to eclipse-like events caused by orbiting circumstellar material. 
Archival Hubble Space Telescope imaging of the region reveals a second, faint (F814W$\sim$20.3 mag) 
companion to HBC 340 that lies 1\farcs02 ($\sim$235 AU) east of the protostar. If associated, 
this probable substellar mass object (20--50 Jupiter masses), HBC 340B, is likely unrelated 
to the observed brightness variations. The sustained brightening of HBC 340 since late 
2014 can be explained by an EXor-like outburst, the recovery from a long duration eclipse 
event caused by obscuring circumstellar dust, or by the gradual removal of extincting material 
from along the line of sight. Our analysis here favors one of the extinction scenarios.
\end{abstract}

\keywords{stars: pre-main sequence; stars: variables: T Tauri, Herbig Ae/Be}

\section{Introduction}

Brightness fluctuations in nebulae associated with pre-main sequence stars are well-documented 
in the literature, most notably by Hubble (1916) who observed changes in the structural details 
of NGC\,2261, the fan-shaped or cometary reflection nebula illuminated by the Herbig Ae/Be star 
R Mon. The variations Hubble observed occurred over timescales of weeks or months, suggesting to 
him that internal motions of nebular material were responsible for the changing illumination pattern. 
The generally accepted explanation today is that streamers of dust forming several AU from R Mon 
cast shadows through an evacuated cavity in the surrounding envelope and onto the walls of the 
parabolic-shaped, reflection nebula (Lightfoot 1989; Close et al. 1997).

Other noteworthy examples of variable nebulae associated with pre-main sequence stars include
NGC\,6729, the R Coronae Australis reflection nebula (Knox Shaw 1916; 1920) and RNO 125, which
is illuminated by PV Cephei lying near the edge of the L1155 and L1158 dark clouds (Cohen et al. 1977; 1981).

More recently, McNeil's Nebula, located $\sim$12' south of NGC\,2068 (M78) in the Orion B 
molecular cloud complex, dramatically increased in brightness during 2003. Lying at the apex 
of the nebula is the deeply embedded protostar IRAS 05436-0007 (V1647 Ori) which brightened by 
approximately five magnitudes from effective obscurity to $r'=$17.7 before fading in early 2006. 
Reipurth \& Aspin (2004) suggested that an EX Lupi (EXor) or FU Ori (FUor) type eruption event 
dispersed a layer of extincting material, permitting the outflow cavity to be illuminated by 
stellar radiation. A subsequent outburst of V1647 Ori occurred in mid-2008 and the source has 
since remained at an elevated photometric state. The structure of the nebula during these peaks 
in brightness in 2003, 2008, and into 2011 show remarkable similarities, supporting the variable 
illumination hypothesis.

In October 2014, three amateur astronomers, Rainer Spaeni, Christian Rusch and Egon Eisenring
working as ``astroteamCERES" from Switzerland noted that the compact reflection nebula associated 
with HBC 340 and HBC 341, two partially embedded young stellar objects in NGC\,1333 and the L1450 
molecular cloud, had changed in appearance, fading significantly from images they had obtained in 
2012. The three astronomers contacted one of the authors (LAH) regarding their discovery and continued 
monitoring the nebula for further brightness variations. 

The Palomar Transient Factory (PTF) has imaged the NGC\,1333 star forming region in a low cadence mode 
since 2011 searching for pre-main sequence outburst phenomena. Inspection of the PTF images confirmed 
the brightness variations in the nebula reported by Spaeni, Rusch and Eisenring, including a distinct 
minimum that occurred in late August and early September 2014. By December 2014, the nebula had returned 
to its previous surface brightness level, as reported by Hillenbrand et al. (2015) in their astronomer's
telegram. It is suggested that the nebula be referred to as the NGC\,1333 Rusch-Eisenring-Spaeni variable nebula 
(or RESVN), after the discoverers of its variable nature. Shown in Figure 1 are two PTF $R-$band images 
centered on HBC 340 obtained near minimum in September 2014 and near peak brightness in January 2015 that 
demonstrate the observed changes. A time-lapse movie of the full complement of PTF images spanning from 
2011 to 2015 is available online at URL: http://www.astro.caltech.edu/\textasciitilde lah/hbc340.

\begin{figure}
\figurenum{1}
\epsscale{1.}
\plottwo{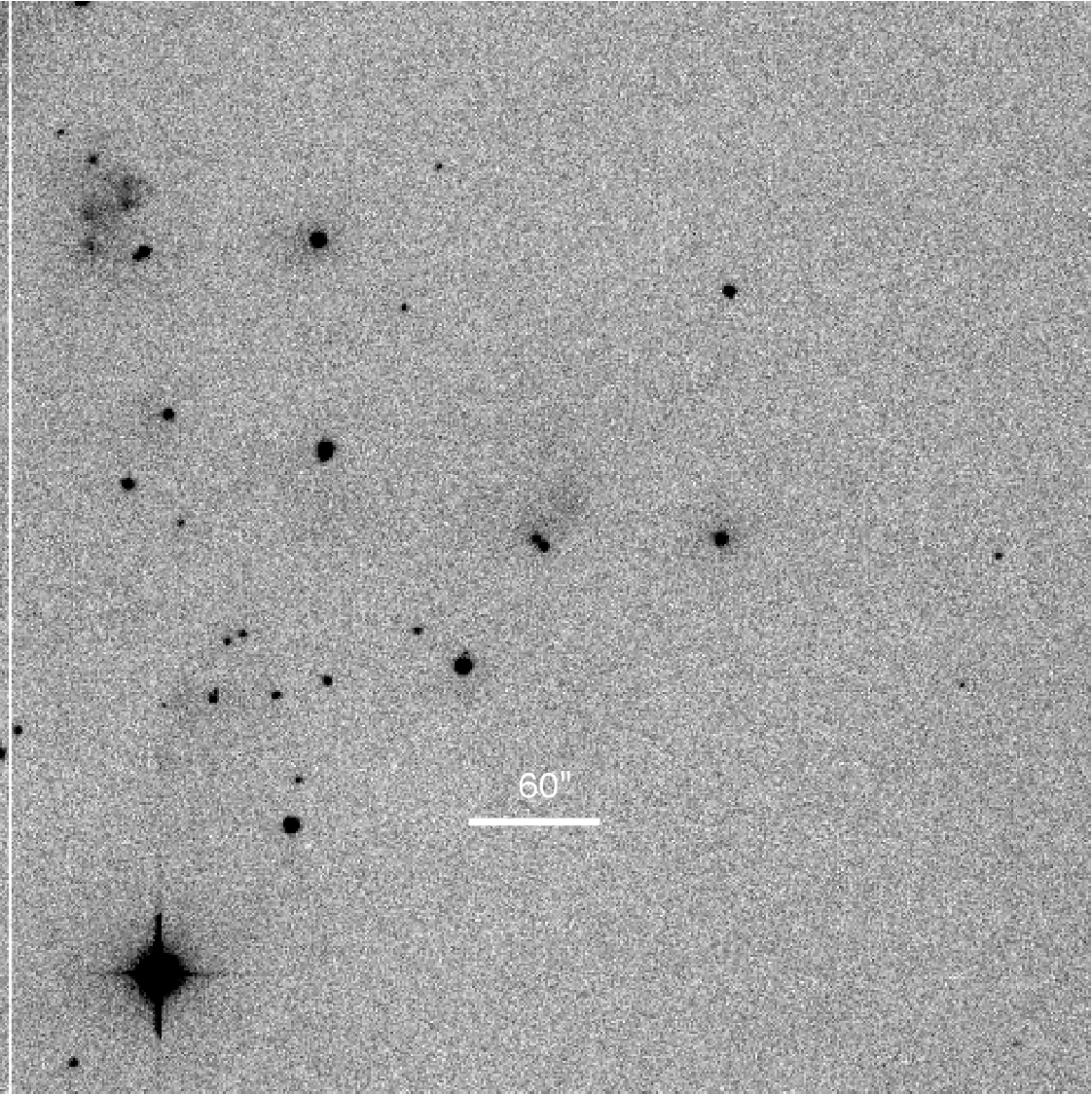}{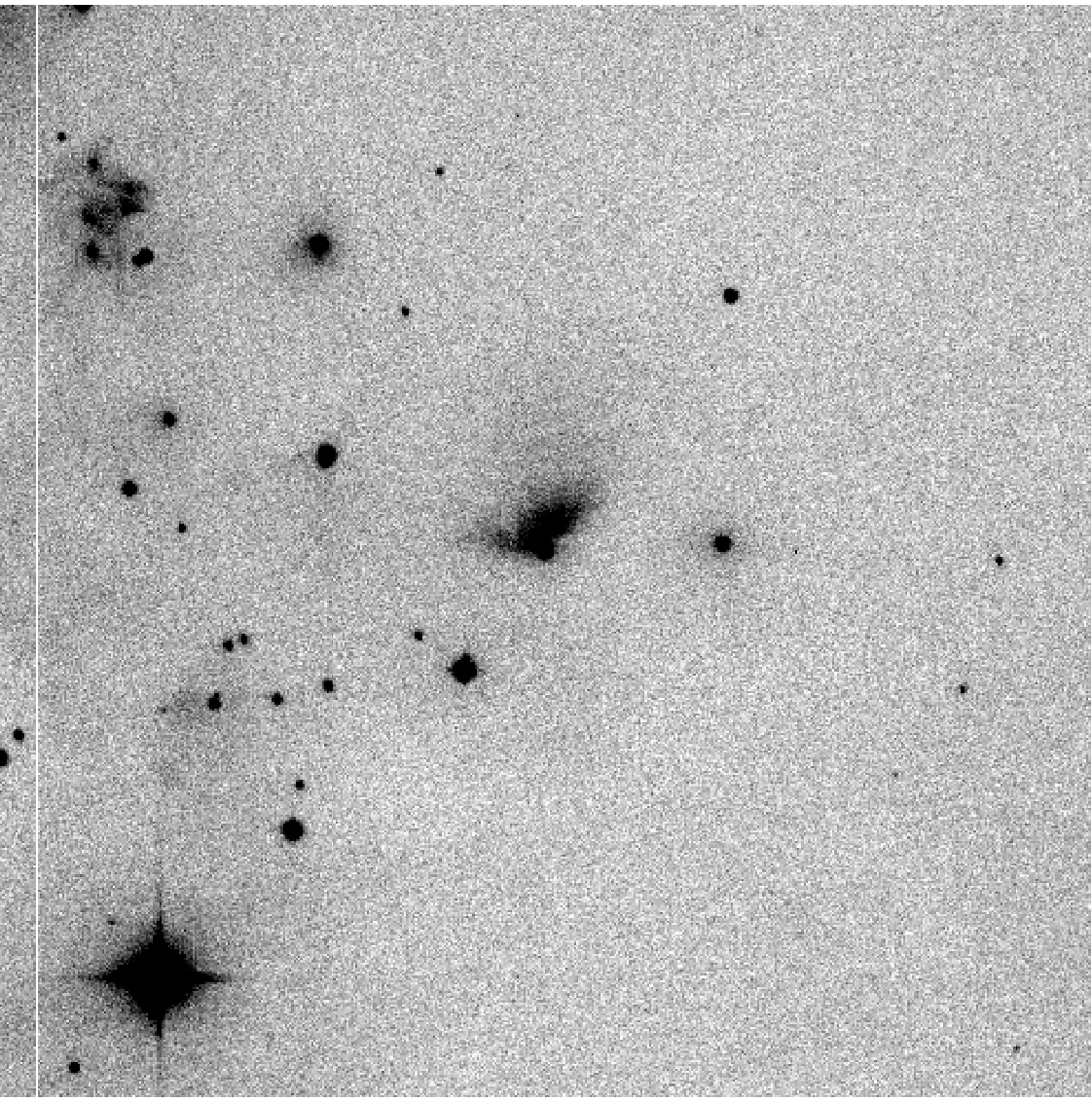}
\caption{Palomar Transient Factory 500$\times$500 arcsec, $R-$band images (north is up, east 
is to the left) of HBC 340 and its associated reflection nebula obtained near minimum light 
in September 2014 (left) and after recovering its former brightness level in January 2015 (right). 
To the northeast lies the complex series of knots of emission that comprise Herbig-Haro (HH) 12, 
one of the more prominent HH objects in the NGC\,1333 star forming region.} 
\end{figure}

The nebular fading appears to be an ongoing phenomenon over the last several decades at least.
Inspection of the red Palomar Observatory Sky Survey (POSS) plates obtained between 1950-1957 
(POSS I) and 1989-1999 (POSS II), suggests that the nebula was brighter during the earlier epoch 
of imaging. Red Lick Observatory plates obtained by Herbig (1974) in 1959 using the 120-inch Shane 
telescope illustrate a state in which the eastern side of the nebula is faint while the western 
side is bright. This is reminiscent of PTF images made when the nebula was either entering or 
recovering from a deep minimum. Optical imaging of the region by one of the authors (SED) in 1999 
using the University of Hawaii (UH) 2.2 m telescope again shows the reflection nebula in a bright 
state. While circumstantial in nature, these few images of the region are suggestive of long-term
variability.

In this contribution, we identify HBC 340, a young protostar located at the apex of the reflection nebula, 
as the primary source of the nebular brightness variations. HBC 340 forms a visual pair with 
the M5-type pre-main sequence star HBC 341, lying $\sim$5\farcs1 to the northeast. Adopting the distance 
of NGC\,1333 derived by Hirota et al. (2007; 2011), $\sim$235$\pm18$ pc, the angular separation 
of HBC 340 and HBC 341 implies a projected physical distance of $\sim$1200 AU, consistent with a 
widely separated, pre-main sequence binary (Mathieu 1994). 
 
HBC 340 was first noted to be variable by Herbig (1974) and was subsequently identified as the bright 
near-infrared source SSV 9A by Strom et al. (1976). Cohen \& Kuhi (1979) spectroscopically classified 
HBC 340 and HBC 341 as K7e and M4.5e, respectively, classifications that were upheld by Winston et al. (2009), 
K7 and M5, in their near-infrared spectroscopic study. A single, bright mid-infrared source was cataloged 
by IRAS and {\it WISE} that is coincident with HBC 340. Using mid-infrared spectral indices determined from 
their {\it Spitzer} Infrared Spectrograph (IRS) 8--30 $\mu$m spectrum of HBC 340, Arnold et al. (2012) 
classified the protostar as a flat spectrum source, i.e. between Class I and Class II, that suffers 
significant visual extinction. Their low-dispersion, IRS spectrum exhibits 10 and 20 $\mu$m silicate 
emission features, similar to other disk-bearing sources.

The observations presented here include PTF imaging of the region and $R-$band photometry of HBC 340; 
one epoch of multiband optical photometry ($BVR_{C}I_{C}$); archival Hubble Space Telescope (HST) 
optical imaging of the nebula; high dispersion optical spectroscopy of HBC 340 and HBC 341 acquired 
on Keck I; and moderate dispersion, near infrared spectroscopy of HBC 340 and HBC 341 obtained using 
SpeX on the NASA Infrared Telescope Facility (IRTF). We discuss the photometric light curve of HBC 340 
produced from the PTF $R-$band imaging, the physical mechanism(s) responsible for its variability and 
current brightening trend, as well as the variable nature of the reflection nebula.

\section{Observations}

Low cadence imaging of NGC\,1333 was obtained using PTF (Rau et al. 2009) and its 7.8 square degree 
mosaic imager coupled to the Samuel Oschin 48-inch Schmidt telescope. Images were obtained in $R-$band, 
typically two frames per night, separated in time by several hours to identify asteroids passing through 
the field of view. The standard integration time is 60 s, yielding a 5$\sigma$ detection limit of 20.5 mag 
in $R-$band. The PTF data are reduced and archived by the Infrared Processing and Analysis Center (IPAC), 
which provides astrometrically and photometrically calibrated images as well as point source catalog
files for each of the regions imaged.

A single epoch of $BVR_{C}I_{C}$-band imaging was obtained on 16 October 1999 using the UH 2.2 m telescope 
on Maunakea with the Tek2K CCD camera installed. The region of interest was serendipitously visible near 
the edge of the 7\farcm5 diameter field of view. The integration times for all frames presented here were 
300 s. The Tek2K images were reduced using standard routines available in the Image Reduction and Analysis 
Facility (IRAF). Aperture photometry was subsequently performed using the {\it phot} task within the
{\it daophot} package of IRAF and calibrated using a Landolt (1992) standard field (SA 113) that was 
observed periodically throughout the night at a range of airmasses.

HST observed the variable nebula using the Advanced Camera for Surveys (ACS) on 31 August 2016 as part
of proposal ID number 14212 (PI Stapelfeldt). The Widefield Camera (WFC) having a platescale of
$\sim$0\farcs05 per pixel was used to image the region in two broadband filters, F606W and F814W.
Integration times for the two F606W frames were 55 and 680 s and for the single F814W frame, 34 s. 
Calibrated frames from the HST archive processed using the On the Fly Reprocessing (OTFR)
system were drizzle-combined with cosmic ray rejection. The two WFC chips were mosaicked together to
create a single image of the region. No further processing was necessary given the limited science
objectives of this program. 

High-dispersion optical spectra of HBC 340 and HBC 341 were obtained on the nights of 27 October 2015,
30 October 2015, and 13 January 2017 (UT) using the High Resolution Echelle Spectrometer (HIRES), 
Vogt et al.\ (1994), on the Keck I telescope. HIRES was configured with the red cross-disperser 
and collimator in beam for all observations presented here. Either the C1 or B1 deckers 
(0\farcs87$\times$7\farcs0 and 0\farcs55$\times$7\farcs0) were used, providing spectral 
resolutions of $\sim$45,000 ($\sim$6.7 km s$^{-1}$) and 66,000 (4.5 km s$^{-1}$), respectively. 
Near-complete wavelength coverage from $\sim$4300 to 8600 \AA\ was achieved. The CCDs were used in 
low-gain mode, resulting in readout noise levels of $\sim$2.8, 3.1, and 3.1 e$^{-1}$ for the red, 
green, and blue chips, respectively. Internal quartz lamps were used for flat fielding and ThAr lamp 
spectra were obtained for wavelength calibration. Integration times ranged from 600 to 1800 s, yielding 
signal-to-noise levels of up to $\sim$100 for the reddest orders. The cross-dispersed spectra were reduced 
and extracted using the MAunaKea Echelle Extraction (MAKEE) pipeline written by T. Barlow. 

Moderate-dispersion near infrared spectra (0.8-2.5 $\mu$m) were obtained for HBC 340 and HBC 341 using 
SpeX on the NASA IRTF on Maunakea. The observations were made on the night of 03 October 2015 (UT) under 
clear conditions and 0\farcs6 seeing. The spectra were obtained in the short cross-dispersed mode (SXD) 
using the 0\farcs3 slit yielding a spectral resolution of $R\sim$2000. Integration times were 120 s per nod 
position for a total of 40 minutes of on source integration time. An A0 V telluric standard within a few 
degrees of and at a similar airmass was observed for each target. Arc lamps and flats were obtained for 
each source to avoid potential calibration problems induced by instrument flexure or changing slit position 
angle. The SpeX spectra were reduced and extracted using SpeXtool, an IDL-based package that provides for 
sky subtraction, flat fielding, wavelength calibration and optimal extraction (Cushing et al. 2004).
Telluric correction was performed using {\it xtellcor}, an IDL routine that interpolates over broad hydrogen 
absorption features using a technique developed by Vacca et al. (2003).

\section{Observational Results}

\subsection{Derived Properties of HBC 340 and HBC 341}

HBC 340 and HBC 341 are embedded within an extensive complex of molecular gas and dust in the NGC\,1333 
star forming region. To estimate their age and mass using pre-main sequence evolutionary models, accurate 
placement on the extinction-corrected, color-magnitude diagram is necessary, requiring knowledge of the 
reddening suffered by both sources from obscuring dust. Shown in the left panel of Figure 2 is the $H-K_{S}$, 
$J-H$ color-color diagram plotted using photometry obtained from the Two-Micron All Sky Survey (2MASS) point 
source catalog. Superposed as dashed lines are the dwarf colors of Pecaut \& Mamajek (2013) and the classical 
T Tauri star locus of Meyer et al. (1997). Both sources lie well to the right of the reddening boundary for 
normal stars, suggestive of circumstellar disk emission. Dereddening both sources to the classical T Tauri 
star locus, we find that HBC 340 and HBC 341 suffer $A_{V}$$\sim$8.0 mag and 3.5 mag of extinction, respectively.

Using these derived extinction estimates and the calibrated photometry from the UH 2.2 m telescope, we plot 
both sources on the $V-I_{C}$, $V$ color-magnitude diagram shown in the right panel of Figure 2. Dotted lines
connect the observed data points with their extinction-corrected counterparts, plotted assuming the intrinsic
colors for pre-main sequence stars taken from Pecaut \& Mamajek (2013). The Siess et al. (2000) evolutionary
models are used to estimate the age and mass of HBC 340 and HBC 341, which are found to be $\sim$0.1 Myr and 
$\sim$1.0 M$_{\odot}$ for the former and $\sim$2.1 Myr and $\sim$0.26 M$_{\odot}$ for the latter. The assigned 
spectral types of HBC 340 and HBC 341 imply effective temperatures of 3970 K and 2880 K, respectively, taken 
from the pre-main sequence temperature scale of Pecaut \& Mamajek (2013).

The above age and mass estimates for HBC 340 and HBC 341 must be regarded with some caution. In addition to 
observational error, the uncertainties associated with these estimates are difficult to quantify, but can be 
divided into two broad categories: those associated with the underlying physics used in modeling stellar 
evolution from the birth line to the zero age main sequence; and the uncertainties induced when transforming 
between the theoretical and observational planes. Hillenbrand \& White (2004) compared the dynamically 
determined masses of 115 main sequence and pre-main sequence stars with their predicted masses from seven 
different theoretical models. They found reasonable agreement for masses above 1.2 M$_{\odot}$, however,
below this value they found discrepancies in mass that ranged from 10--30\%. Over all age ranges, 
Hillenbrand \& White (2004) concluded that systematic discrepancies between the dynamically determined masses 
and the track-predicted masses were dominated by errors in the theoretical treatment of convection and opacity. 

The transformation between the theoretical and observational planes is typically accomplished by fitting main sequence 
colors and bolometric corrections as a function of effective temperature. The intrinsic colors and bolometric 
corrections of pre-main sequence objects remain poorly established. Uncertainties in distance, reddening (critical 
for the heavily extincted HBC 340), and binarity further complicate the placement of pre-main sequence objects on 
the color-magnitude diagram. Table 1 summarizes the observed and derived properties of both young stellar objects 
with uncertainties either determined or estimated. 

\begin{figure}
\figurenum{2}
\includegraphics[angle=90, height=6cm]{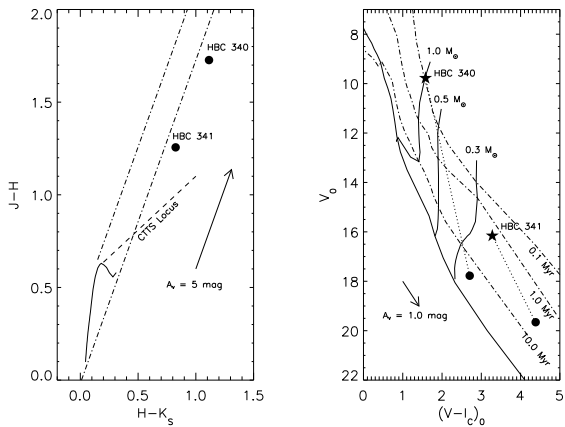}
\caption{The $H-K_{S}$, $J-H$ color-color diagram (left panel) of HBC 340 and HBC 341 plotted using 2MASS
photometry. Overplotted are the main sequence colors of Pecaut \& Mamajek (2013), solid line, and the
classical T Tauri star locus of Meyer et al. (1997), dashed line. The approximate reddening boundaries for
normal stars (dot-dashed line) are shown with slopes derived using the extinction coefficients for diffuse interstellar clouds
taken from Martin \& Whittet (1990). De-reddening the sources to the CTTS locus, we estimate extinctions of 
$A_{V}$$\sim$8.0 mag and 3.5 mag for HBC 340 and HBC 341, respectively. The $V-I_{C}$, $V$ color-magnitude 
diagram is shown in the right panel with the observed (solid points) and extinction-corrected (five-pointed 
stars) positions for each source connected by dotted lines. The evolutionary tracks and isochrones of the Siess 
et al. (2000) models are superposed.}
\end{figure}

\subsection{Optical Imaging of the Reflection Nebula}

Visual inspection of Figure 3, a deep $VR_{C}I_{C}$-band composite image of the region obtained in 1999
using the UH 2.2 m telescope, and Figure 4, a composite $izy$-band image from the Panoramic Survey
Telescope and Rapid Response System (PANSTARRS), suggests that the variable reflection nebula is an 
illuminated wall of a cavity carved out of the enveloping molecular cloud, presumably by an outflow 
emerging from HBC 340. The remnant molecular material appears distinctly toroidal in shape, with the
dark lane visible below HBC 340 presumably lying in the foreground. The opening in the north reveals 
the brightly illuminated ridges of the cavity interior which remain faintly visible, even when the 
nebula is near minimum (Figure 1). Directly south of HBC 340, the edge of the obscuring molecular 
material appears to be illuminated, possibly the bipolar symmetrical counterpart of the variable 
nebula to the north.

\begin{figure}
\figurenum{3}
\includegraphics[angle=0,height=4.6cm]{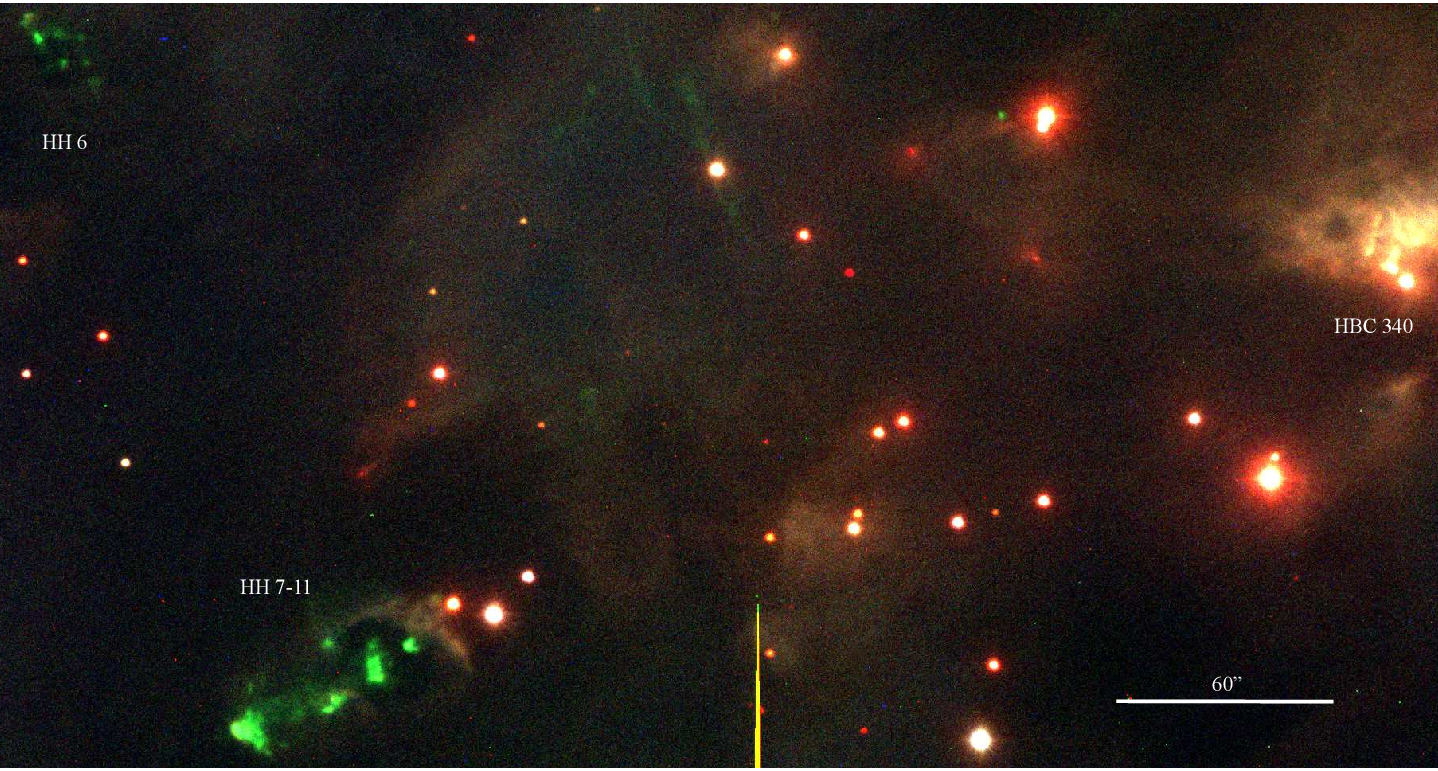}
\caption{Composite, three-color $VR_{C}I_{C}$ image of the HBC 340 region in NGC\,1333 obtained using the 
Tek2K CCD camera on the UH 2.2 m telescope in October 1999 oriented such that north is up, east is to the 
left. The field presented here spans several arcminutes from HBC 340 and the variable reflection nebula 
near the right edge, to the impressive series of HH objects 7-11 on the bottom left. HH 6 lies near the
top, left edge of the frame. The image suggests that extensive molecular material lies in the foreground 
of HBC 340 and HBC 341, but that an opening to the north reveals the illuminated interior of the evacuated 
cavity.}
\end{figure}

\begin{figure}
\figurenum{4}
\includegraphics[height=8.75cm, angle=270]{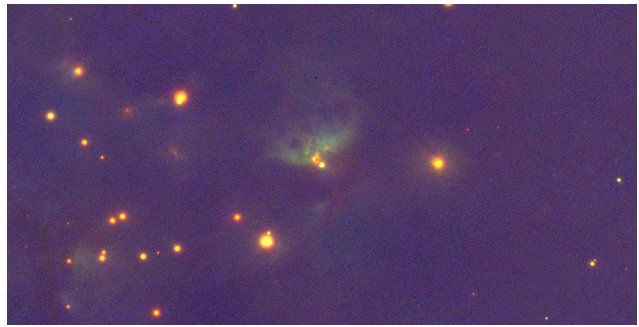}
\caption{Composite, three-color $izy-$band image of the HBC 340 region obtained from the PANSTARRS image 
server oriented such that north is up and east is to the left. The field of view presented here is about
5\farcm9$\times$3\farcm0 in dimension. The remnant molecular material surrounding HBC 340 and HBC 341 appears 
toroidal in shape, as if a bipolar outflow emerging from the protostar has carved out the visible cavity.}
\end{figure}

The archival HST ACS imaging of the reflection nebula reveals a faint companion to HBC 340, 1\farcs02 east
of the protostar. Assuming the distance derived by Hirota et al. (2007; 2011), this angle implies 
a projected physical separation of $\sim$235 AU. From aperture photometry of the source, we determine 
$m_{F814W}$=20.33$\pm$0.03 mag. Given the heavy extinction in the region and the close proximity of the 
source to HBC 340, we assume it to be a physical companion of the protostar. Using the extinction values 
derived for HBC 340 ($A_{V}$=8.0 mag) and HBC 341 ($A_{V}$=3.5 mag), we estimate its absolute magnitude 
to be between $M_{F814W}$=9.6 and 11.8 mag. Adopting the pre-main sequence models of Baraffe et al. (1998) 
for an age of $\sim$1.0 Myr, we estimate the mass of the companion to be between $\sim$20 and 50 times that 
of Jupiter, i.e. a brown dwarf. The dynamical timescale of its orbit suggests that HBC 340B plays no role in 
the observed brightness variations of the reflection nebula.

Shown in Figure 5 is a 25"$\times$25" subsection of the deep ACS F606W image of the region, centered
within the reflection nebula. HBC 340B is readily apparent as is a bright rim of molecular material
northeast of HBC 341 that is likely illuminated by HBC 340. South of the protostar, the dark edge
of the foreground molecular cloud evident in Figures 3 and 4 stands out prominently. The ACS image reinforces
the supposition that HBC 340 lies within a conically-shaped pocket that has been cleared of substantial
molecular gas and dust. The PSF of HBC 340 is saturated in this frame and in the single F814W image,
but that of HBC 341 reveals a stellar, pointlike source. The shallow 55 s F606W image is less impacted
by charge blooming around HBC 340 permitting a more thorough examination of its PSF. Some structure is
apparent that distinguishes it from an unresolved point source, particularly to the south where the
obscuring foreground cloud may have an impact. There is tentative evidence that the PSF of HBC 340
is elongated, but charge blooming prevents a definitive conclusion.

\begin{figure}
\figurenum{5}
\includegraphics[angle=0, height=12cm]{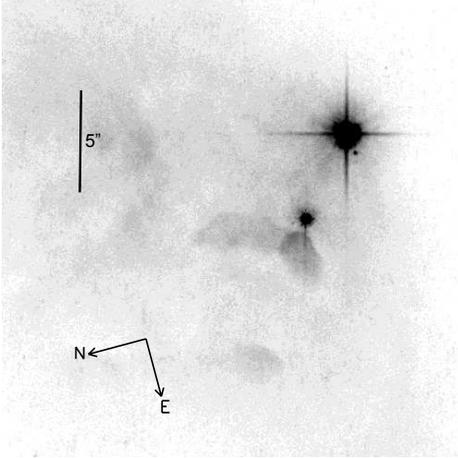}
\caption{A 25"$\times$25" subsection of the deep ACS WFC F606W image of the reflection
nebula and HBC 341 to the northeast and HBC 340 to the southwest. A much fainter source 
HBC 340B is prominent in the image lying 1\farcs02 east of the protostar. If associated 
with HBC 340, this source is estimated to have a mass between 20 and 50 times that of 
Jupiter, as determined from the pre-main sequence models of Baraffe et al. (1998).}
\end{figure}

Analogous to NGC\,2261 and McNeil's nebula, we attribute the brightness fluctuations in the reflection 
nebula to variable illumination originating from HBC 340 positioned near its apex. From the PTF image archive, 
we obtained 124 frames approximately 300\arcsec$\times$300\arcsec\ in dimension and corrected them for 
positional registration and flux differences using the mean offset of sky-subtracted photometry for five 
bright stars in the field. These comparison stars range in brightness from $R=$14.5 to 16.5 mag and exhibit 
a measured scatter of 0.01 to 0.07 mag, consistent with the level of variability expected in quiescent pre-main 
sequence stars. Photometry of HBC 341, the fainter of the two sources, is contaminated by the brighter HBC 340, 
but the extracted light curves reveal that HBC 340 is the more dramatic variable of the two with changes of 
a magnitude or more occurring over timescales of weeks to months. Shown in Figure 6 is the PTF $R-$band light-curve 
of HBC 340 spanning over six years from January 2011 to February 2017. Since the deep minimum that occurred in 
September 2014, HBC 340 has brightened by more than two magnitudes. The two nebular fadings captured by PTF 
that occurred near Julian dates (JD) 2456630 and 2456900 correspond to two local minima in the light curve of 
HBC 340. From this we conclude that HBC 340 is the dominant source of the observed nebular variations. 

\begin{figure}
\figurenum{6}
\includegraphics[angle=90, height=6cm]{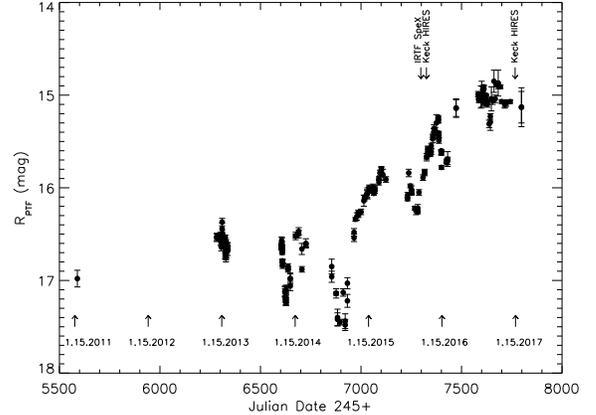}
\caption{The PTF $R-$band light curve of HBC 340 plotted as a function of Julian date. The figure spans
over six years of low-cadence observations made of the region. HBC 340 has brightened by more than two 
magnitudes since a deep minimum that occurred in late August and early September 2014. Several local 
minima are evident in the light curve, which correspond to faint states of the diffuse nebula.}
\end{figure}

In addition to Herbig's (1974) comment, the limited photometry of HBC 340 available from the literature 
is suggestive of variability. Cohen's (1980) $V-$band magnitude (17.7) of HBC 340 is consistent with 
that determined from the 1999 UH 2.2 m observations, $V=$17.78. The USNO B-1.0 catalog provides an 
$R-$mag of 15.89 for HBC 340, but the photometric uncertainties associated with the catalog are large, 
$\sim$0.25 mag. The 1999 UH 2.2 m photometry found $R_{C}$=16.44$\pm$0.004 mag, which is comparable to 
the protostar's apparent baseline before the onset of the long-term, brightening trend in late 2014.
PANSTARRS finds the minimum and maximum $r-$band magnitude of HBC 340 to range between 16.728 and 
17.603 mag over 21 single epoch observations. Variation of similar or lesser amplitude is observed in 
the PANSTARRS $i, z,$ and $y-$band imaging as well.

In the near infrared, Aspin et al. (1994) published magnitudes for HBC 340 of: $J=$13.21, $H=$11.40, and 
$K=$10.19, which differ substantially from those determined by 2MASS: $J=$12.588, $H=$10.861, and 
$K_{S}$=9.747. In the thermal and mid-infrared, inspection of the {\it Spitzer} IRS spectrum of HBC 340 
obtained by Arnold et al. (2012) shows that the observed flux level is a factor of two lower than the 
IRAC and MIPS photometry of Gutermuth et al. (2008). Additionally, WISE photometry of HBC 340 
(3.37 $\mu$m=8.422$\pm$0.22 mag; 4.62 $\mu$m=7.198$\pm$0.018 mag; 12.08 $\mu$m=4.566$\pm$0.016 mag; 
22.19 $\mu$m=1.691$\pm$0.021 mag) is fainter at $L-$ and $M-$band than measured by {\it Spitzer} 
(IRAC 3.55 $\mu$m=7.82$\pm$0.01 mag; 4.49 $\mu$m=6.86$\pm$0.01 mag; 5.73 $\mu$m=5.91$\pm$0.01 mag; 
7.87 $\mu$m=4.90$\pm$0.01 mag; MIPS 24 $\mu$m=1.32$\pm$0.01), even though the source is likely confused 
with HBC 341 in all bands of the WISE photometry. Taken together these limited observations of HBC 340 
are suggestive of variability, which extends into the thermal and mid-infrared.

\subsection{Keck High Dispersion Spectroscopy of HBC 340 and HBC 341}

The origin of periodic variability in pre-main sequence stars is generally attributed to spot-induced 
rotational modulation, while irregular variability arises from accretion, or obscuration from dust within 
the circumstellar environment, or stellar flare events (e.g. Cody et al. 2014; Stauffer et al. 2016). 
Classic signatures of accretion activity include strong H$\alpha$ emission (Bertout 1989), Paschen and 
Brackett series emission (Muzerolle et al. 1998), forbidden line emission (Edwards et al. 1987; Simon et al. 2016), 
and \ion{He}{1} and \ion{Ca}{2} emission in the optical and near infrared (Hamann \& Persson 1992). 
The Keck high dispersion spectra of HBC 340 and HBC 341 reveal strong H$\alpha$ emission in both sources, 
low excitation forbidden emission lines of [O I], [N II], and [S II], that are generally associated with 
collimated jets or disk winds, and emission core reversal in the profiles of the \ion{Ca}{2} near infrared 
triplet. Shown in Figure 7 are the H$\alpha$, [S II] $\lambda\lambda$6717, 6731 and \ion{Ca}{2} $\lambda$8542 
emission line profiles of HBC 340 in 2015 and 2017 and of HBC 341 in 2015. In Table 2, we provide the measured 
equivalent widths of select emission features present in both sources. The strengths of the emission features 
found in the spectra of HBC 340 and HBC 341 are indicative of accretion. Of particular interest is that although 
the HIRES spectra of HBC 340 were obtained 15 months apart during which the protostar brightened by nearly a 
magnitude at $R-$band, the January 2017 spectrum exhibits significantly weaker H$\alpha$ emission (factor of two), 
no detectable H$\beta$ emission, and forbidden emission lines with equivalent widths that are about half of their 
October 2015 strengths. 

\begin{figure}
\figurenum{7}
\includegraphics[angle=90, height=6cm]{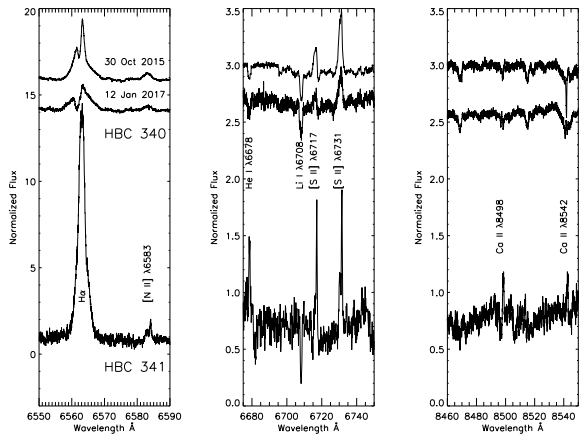}
\caption{Select orders of the HIRES spectra of HBC 340 obtained on 30 October 2015 and 
12 January 2017 and of HBC 341 obtained on 30 October 2015 centered near H$\alpha$ (left panel), 
[S II] $\lambda\lambda$6717, 6731 (center panel) and \ion{Ca}{2} $\lambda$8542 (right panel). 
The strong emission features present are indicative of disk accretion in both sources.}
\end{figure}

The optically thin, forbidden emission lines of [O I] $\lambda\lambda$6300, 6363, [S II] $\lambda\lambda$6717, 
6731 and [N II] $\lambda\lambda$6548, 6583 evident in the spectra of HBC 340 are blueshifted on average by 
$\sim$25 km s$^{-1}$ relative to the protostar, suggesting that these profiles are dominated by a low velocity
component (LVC) generally attributed to slow disk winds (Simon et al. 2016). A higher velocity component is 
present in the [O I] $\lambda$6300 emission line that is blueshifted by $\sim$68 km s$^{-1}$ relative to the 
protostar, which is thought to arise from an emerging jet. The spectrum of HBC 341 
shows two distinct components of [N II] $\lambda$6583 and [S II] $\lambda\lambda$6717, 6731 emission: a higher 
velocity component that is blueshifted by $\sim$45 km s$^{-1}$ and a LVC that is red-shifted by $\sim$15 km s$^{-1}$ 
relative to the star. Both sources appear to be associated with disk winds as well as accretion induced jets.

Under the assumption that magnetospheric accretion is responsible for the observed hydrogen line emission in 
HBC 340 (e.g. Muzerolle et al. 2001; Kurosawa et al. 2006), the mass accretion rate can be estimated to first 
order by converting the measured H$\alpha$ equivalent widths into line luminosities using the following:

$L_{H\alpha} = 4 \pi d^{2} W_{H\alpha} \frac{F6563}{F6410} F(6410,0.0) 10^{-0.4 R_{C}} (1 + r_{\lambda})$ 

\noindent where F(6410,0.0) represents the flux of a zeroth magnitude star in the $R_{C}$ filter using 
wavelength units and $r_{\lambda}$ is the veiling suffered at this wavelength (assumed here to be negligible 
given the weak emission present). We estimate the extinction corrected, R$_{C}$ magnitude by assuming 
A$_{V}\sim$8.0 mag and using the normal extinction law ($R=3.1$, Mathis 1990) to derive A$_{R}\sim$6.0 mag. 
Adopting a distance, $d$, of 235 pc (Hirota et al. 2007; 2011) and assuming the ratio of fluxes at $\lambda$6563 
and $\lambda$6410 to be near unity, we calculate the accretion luminosity using the following relationship 
from Alcala et al. (2014):

$ log(\frac{L_{acc}}{L_{\odot}}) = (1.50\pm0.26) + (1.12\pm0.07) log(\frac{L_{H\alpha}}{L_{\odot}})$.

\noindent From the accretion luminosity, the mass accretion rate, $\dot{M}$, follows from Gullbring et al. (1998):

$ \dot{M} = \frac{L_{acc} R_{*}}{G M_{*} (1 - R_{*}/R_{in})}$ 

\noindent where R$_{in}$ is the inner disk radius and assumed to be 5 R$_{*}$, M$_{*}$$\sim$1.0 M$_{\odot}$ 
(from the pre-main sequence models of Siess et al. 2000) and R$_{*}$$\sim$2.0 R$_{\odot}$. A mass accretion 
rate of $\dot{M}$ = 2.8$\times$10$^{-9}$ M$_{\odot}$ yr$^{-1}$ follows. A factor of two uncertainty 
can be assigned to this value of $\dot{M}$, estimated by propagating the known distance error, 
the stated uncertainties in the accretion luminosity relationship, and the assumed uncertainties 
in the mass and radius of the protostar. The mass accretion rate would be lower still during the
January 2017 HIRES observations given the substantially weaker $W_{H\alpha}$ measurement.

The heliocentric radial velocities of HBC 340 and HBC 341 were determined by cross correlating select orders 
of their spectra that are free of telluric features and emission lines with those of established radial velocity 
standards of similar spectral type using the IRAF task {\it fxcor}. The heliocentric radial velocity of HBC 341 
was determined to be 16.43$\pm$0.86 km s$^{-1}$ on 27 October 2015 and 15.60$\pm$0.74 km s$^{-1}$ on 30 October
2015, which are consistent with each other, but somewhat displaced from the measurement of Foster et al. (2015) 
and Cottaar et al. (2015) in their near infrared, spectroscopic studies of NGC\,1333 and IC\,348: 13.16$\pm$0.46 km s$^{-1}$. 

The heliocentric radial velocity of HBC 340, the brighter of the two sources, was determined to be 17.33$\pm$3.85 km s$^{-1}$ 
on 27 October 2015, 18.17$\pm$5.35 km s$^{-1}$ on 30 October 2015, and 14.76$\pm$2.93 km s$^{-1}$ in January 2017, 
which are consistent with the range of velocities determined by Foster et al. (2015) and Cottaar et al. (2015), 
that span from 15.46$\pm$0.99 to 18.19 $\pm$2.80 km s$^{-1}$. Both our optical measurements and the published
infrared measurements have significantly higher errors associated with the radial velocities measured for the 
brighter HBC 340 than for the fainter HBC 341.

Foster et al. (2015) measure an average $v$sin$i$ for HBC 340 of 58.26$\pm$1.26 km s$^{-1}$, determined from
four measurements made over the course of $\sim$113 days during 2013-2014. Cross-correlating the high signal to
noise HIRES spectrum of HBC 340 from 30 October 2015 with an artificially broadened, template spectrum of a 
slowly rotating, K4-type radial velocity standard, we find the best fitting $v$sin$i$ based upon $\chi^{2}$ 
analysis to be $\sim$13.0$\pm$2.0 km s$^{-1}$. The absorption line widths, however, are arguably better matched 
by an even lower $v$sin$i$ of 5--7 km s$^{-1}$, suggesting that optical veiling has filled in the absorption
profiles to some degree. Shown in Figure 8 is one order of the HIRES spectrum of
HBC 340 plotted with the spun-up template spectra having $v$sin$i$ values of 5, 9, 15, and 20 km s$^{-1}$, which 
clearly demonstrates that HBC 340 is not a rapid rotator.

\begin{figure}
\figurenum{8}
\epsscale{0.75}
\plotone{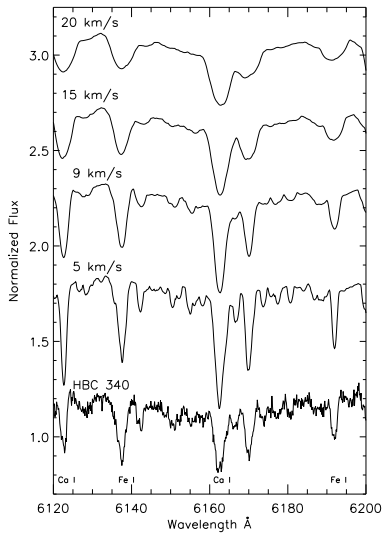}
\caption{One order of the HIRES spectrum of HBC 340 plotted with artificially broadened, template spectra
of a slowly rotating, K4-type radial velocity standard having $v$sin$i$ values of 5, 9, 15, and 20 km s$^{-1}$.
The best fitting $v$sin$i$ of HBC 340 based on $\chi^{2}$ analysis is $\sim$13.0$\pm$2.0 km s$^{-1}$, but this
is heavily influenced by the veiling which also acts to produce smaller line depths; the line widths
are well-matched by an even lower $v$sin$i$, $\sim$5--7km s$^{-1}$.}
\end{figure}

The cross-correlation function of HBC 340 with the slowly rotating, K-type radial velocity standard, however, 
is significantly broadened in appearance, having a full width at half maximum that is two to three times 
that expected for a single, slowly rotating star. We suggest that the broadened cross-correlation function 
is not the result of rapid rotation, but rather the presence of an unresolved, spectroscopic binary companion.
Under this hypothesis, we deconvolve the broadened cross-correlation function into two gaussian profiles and 
recover distinct radial velocities for all three epochs of HIRES observations. The radial velocities of the 
positive and negative components, their associated uncertainties, as well as the systemic velocities, are 
provided in Table 3. Continued high dispersion, spectroscopic monitoring of HBC 340, particularly in the near 
infrared where the absorption features of a late-type companion could potentially be isolated, are needed to 
further refine the radial velocity curves of the two components. 

\subsection{Near Infrared Spectroscopy of HBC 340 and HBC 341}

The IRTF SpeX spectra of HBC 340 and HBC 341 were used to confirm the spectral classifications of the 
young stellar objects by comparing their absorption line strengths at $H-$ and $K-$bands where they
are least impacted by extinction, with those of dwarfs and giants taken from the IRTF spectral library 
(Cushing et al. 2005; Rayner et al. 2009). The telluric-corrected, near-infrared spectrum of HBC 340 
shown in Figure 9 exhibits strong \ion{He}{1} $\lambda$10830 emission as well as Pa$\beta$, \ion{Fe}{2}, 
Pa$\alpha$, and Br$\gamma$ emission, suggestive of accretion activity. The equivalent widths of these 
emission features are provided in Table 2. Also present in the spectrum of HBC 340 are weak H$_{2}$ 
emission lines at 1.958, 2.122 and 2.406 $\mu$m, implying the presence of shocked molecular gas, potentially 
resulting from the impact of jets or outflows into molecular material. The H$_{2}$ $\nu$=1-0 S(1) 2.122 $\mu$m 
emission line probes the inner disk region where dust is optically thick and requires a high column density 
to become visible over the near infrared continuum of the protostar. The CO bandheads of HBC 340 are in 
absorption, intermediate in strength between those of a dwarf and a giant, suggestive of low surface gravity 
and a sub-giant luminosity class. The stellar spectral energy distribution (SED) flattens considerably in 
this region, implying infrared excess arising from circumstellar disk emission. 

Given the significant extinction suffered by HBC 340, we can also estimate the mass accretion rate using the
Pa$\beta$ and Br$\gamma$ line luminosities where extinction is reduced by factors of 3.0 and 6.8, respectively,
relative to H$\alpha$. Using the equivalent widths presented in Table 2, we determine the accretion luminosity
from equations (1) and (2) of Muzerolle et al. (1998) and derive mass accretion rates of $\dot{M}$ = 5.6$\times$10$^{-9}$
M$_{\odot}$ yr$^{-1}$ for Pa$\beta$ and $\dot{M}$ = 5.9$\times$10$^{-9}$ M$_{\odot}$ yr$^{-1}$ for Br$\gamma$,
consistent with the value derived using the H$\alpha$ line strength.

While the near infrared spectrum of HBC 340 exhibits emission lines indicative of disk accretion, that of 
HBC 341 is devoid of such features and is suggestive of a heavily extincted, M-type pre-main sequence star. 
The SED of the star turns over beyond $\sim$2.3 $\mu$m, but its slope is clearly distinct from that of a 
pure M5 dwarf photosphere. From this and its near-infrared colors, we conclude that HBC 341 is also associated 
with a circumstellar disk. 

\begin{figure}
\figurenum{9}
\includegraphics[angle=90, height=6cm]{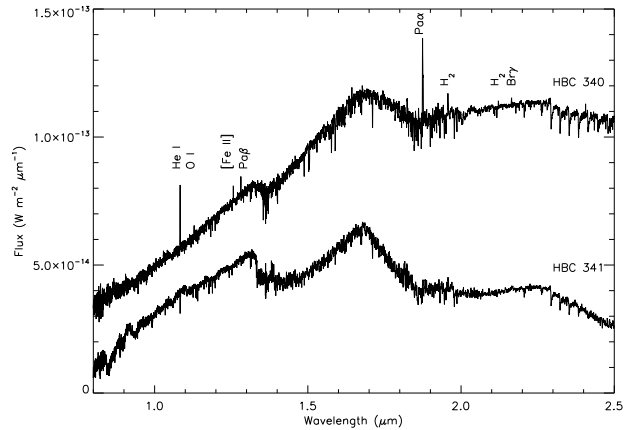}
\caption{Telluric corrected, merged order SpeX spectra of HBC 340 and HBC 341. The spectrum of HBC 340 is 
offset by 2$\times$10$^{-14}$ W m$^{-2}$ $\mu$m$^{-1}$ to separate it from the spectrum of HBC 341 at blue 
wavelengths. Emission features characteristic of circumstellar disk accretion (He I $\lambda$10830, Pa$\beta$, 
Br$\gamma$) are present in the spectrum of HBC 340, but not in that of HBC 341. The slopes of both SEDs 
beyond two microns, however, are suggestive of excess infrared emission, indicative of circumstellar disks.} 
\end{figure}

\section{Discussion}

The broadband spectral energy distribution and {\it Spitzer} IRS spectra demonstrate that HBC 340 is a
heavily extincted, transitional Class I / Class II protostar (Arnold et al. 2012). The optical and near 
infrared spectra of HBC 340 imply that active accretion is occurring from the envelope or circumstellar 
disk, which potentially contributes to its irregular variability on short timescales. Casual inspection 
of the protostar's light curve, however, is suggestive of a correlated signal in the distinct drops of 
varying depth against the general increase in brightness observed over the past several years. Given the 
unevenly sampled light curve, we use the Lomb normalized periodogram and the box fitting least squares 
algorithm to search for evidence of a periodic signal. A high significance level, broad signal is found
centered near $\sim$295 days, corresponding to a Keplerian orbit semi-major axis of $\sim$0.9 AU for a
solar mass protostar. Deviant points are evident in the resulting phase-magnitude diagram for this assumed 
period, but the shape of the dips in the light curve are suggestive of eclipse-like events. The duration 
of these eclipses, $\sim$100 days, relative to the inferred period imply that the proposed spectroscopic 
binary companion, HBC 340Ab, cannot be directly responsible, since the eclipse length would be one to two 
orders of magnitude shorter in duration. HBC 340Ab could, however, play an indirect role by partially 
disrupting or shepherding the circumstellar disk, which then periodically occults the primary. 

The mass ratio of HBC 340Aa to HBC 340Ab can be estimated using the radial velocities determined from the 
highest signal to noise HIRES spectrum from 30 October 2015. Given the fitting procedure used for the 
gaussian profiles on the broad cross correlation function, the radial velocity components are likely to be 
biased downward, i.e. the true velocity amplitude is probably higher than the measurements would indicate. 
With this caveat in mind and assuming the orbital eccentricity to be $\ll$1, we determine the mass ratio
to be near unity using the radial velocities presented in Table 3 and the fundamental relationship:

$\frac{M_{1}}{M_{2}} = \frac{Vr_{2} - \gamma}{\gamma - Vr_{1}} \sim 1.0$

\noindent where $\gamma$ is the systemic velocity of the binary. Accounting for a decrease in brightness 
of $\sim$0.75 mag resulting from the presence of an equal mass companion, we find little impact to the 
assigned mass of HBC 340Aa given the near vertical slope of the solar mass evolutionary track shown in
the right panel of Figure 2. From Kepler's third law, the sum of the stellar masses can be expressed as a 
function of orbital period and the orbital velocities: 

$M_{1} + M_{2} = \frac{P}{2\pi G} \frac{(v_{1}+v_{2})^3}{sin^{3}(i)}$.

\noindent Solving for the period, $P$, where $i$ is the unknown inclination angle, and G is the 
gravitational constant, we find:

$\frac{P}{sin^{3}(i)} \sim$3 years.

The resulting orbital timescale of HBC 340Ab corresponds to a minimum semi-major axis of $\sim$2.6 AU. It 
is possible that a partially disrupted or flared disk rim could lead to the dipping behavior of HBC 340, 
thereby inducing the observed brightness fluctuations visible across the walls of the reflection 
nebula. The {\it Spitzer} IRS spectrum of HBC 340, however, is not suggestive of a cleared inner disk 
analogous to CoKu Tau 4, the pre-main sequence binary in Taurus-Auriga resolved by aperture masking interferometry 
(Ireland \& Kraus 2008; Nagel et al. 2010).

To explain the sustained rise in brightness of HBC 340 since September 2014, an extended feature within a near 
edge-on circumstellar disk could be invoked, similar to KH-15D in NGC\,2264 (Hamilton et al. 2001) or LRLL 35 
in IC\,348 (Cohen et al. 2003). Such a feature would be more broadly extended than that responsible for the 
narrow drops observed against the general rise in brightness. The sparse optical photometry available from the 
literature, however, does not provide evidence for a period when the protostar was at its current bright state. 
It is also conceivable that extincting material is being gradually removed from along the line of sight to the 
protostar, thus causing the slow rise in brightness. Multi-color photometry during the upcoming observing season 
could test this hypothesis.

Alternatively, the variability of HBC 340 and its rise in brightness since September 2014 could be attributed 
to a slow outburst-like event. The two principal classes of pre-main sequence stars that experience 
eruptive outbursts in luminosity are FU Ori and EXor type variables. FU Ori phenomena are characterized by 
a single event during which the star brightens by several magnitudes over the course of several months to 
years and remains at an elevated state for decades or longer. Following outburst, FU Ori type variables exhibit 
a complex, broad-lined absorption spectrum, resembling an F- or G-type supergiant, with P Cygni-like absorption 
structure associated with its H$\alpha$ emission profile, suggestive of strong winds (Herbig 1977). Classical 
examples of FU Ori stars include the prototype FU Ori, V1057 Cyg, and V1515 Cyg.

EXors are eruptive variables that occasionally flare up from minimum light by a magnitude or more over
the course of a few weeks to months before fading once again. During outburst, the class prototype, EX Lupi, 
exhibits a hot emission spectrum that dominates the M-dwarf absorption spectrum present during quiescence
(Herbig 2007). The outbursts presumably result from the rapid infall of circumstellar material, as evidenced 
by inverse P Cygni-like H$\alpha$ emission profiles and optical veiling. Examples of EXor candidates include 
EX Lupi, NY Ori, V1118 Ori and V1143 Ori.

The lightcurve of HBC 340 is reminiscent of an EXor-like variable, and its steady rise in brightness is 
similar to a trend noted by McLaughlin (1946) in the light curve of EX Lupi between 1936 and 1938 when 
the star brightened by more than a magnitude before once again returning to minimum light. McLaughlin (1946)
found that EX Lupi's variations were irregular in nature and that it remained inactive at minimum light 
for years before brightening by $\sim$2.5 magnitudes or more. Conspicuous `nova-like' maxima were noted in
1901, 1914, 1925, 1929, and 1934 that were followed by smaller, irregular fluctuations that persisted for
a year or longer (McLaughlin 1946; Herbig 2001; 2007). In 1955 an extreme outburst occurred during which 
EX Lupi brightened by five magnitudes. Such an event occurred again in early 2008 when the star brightened 
by more than five magnitudes over the course of seven months and was studied extensively by Aspin et al. (2010) 
using high dispersion optical and near infrared spectroscopy. Whether the physical mechanism responsible for 
these extreme outbursts is identical to that of the low amplitude outbursts is unknown.

While the photometric characteristics of HBC 340 are arguably consistent with those of an EXor-like variable, 
spectroscopic comparisons are not quite as parallel. HIRES spectroscopy of HBC 340 does not exhibit the complex 
emission line spectrum superimposed over the late-type continuum evident in EX Lupi during quiescence or in
low amplitude outburst (Herbig 2007). Nor does HBC 340 resemble spectra of EX Lupi obtained during its 2008
extreme outburst, which show strong multi-component \ion{Fe}{1}, \ion{Fe}{2}, and \ion{He}{1} emission, as 
well as complex H$\beta$ and \ion{Na}{1} D emission line profiles (Aspin et al. 2010). Blue-shifted absorption 
features were also observed in the \ion{H}{1}, \ion{Na}{1} and \ion{Ca}{2} profiles of the EX Lupi spectrum, 
which changed in structure as the outburst progressed in early 2008 (Aspin et al. 2010). 

HBC 340 exhibits substantially weaker H$\alpha$ emission in all HIRES observations, W(H$\alpha$)$\sim$10--13\AA,
than reported for EX Lupi by Herbig et al. (2001); W(H$\alpha$)$\sim$30--60\AA\ in quiescence and during low 
amplitude outbursts. The mass accretion rate of HBC 340 is estimated to be two orders of magnitude lower than 
that estimated for EX Lupi during its 2008 extreme outburst, but one order of magnitude greater than that 
estimated for EX Lupi during quiescence by Sipos et al. (2009), 4.0$\times$10$^{-10}$ M$_{\odot}$ yr$^{-1}$.
The low mass accretion rate of HBC 340 could potentially be explained by a cleared inner disk, swept up
by the dynamical interaction of HBC 340Aa and HBC 340Ab. The observed decrease in H$\alpha$ emission line 
strength between 2015 and 2017 would require either a reduction in mass accretion rate (factor of two) or 
a substantial decrease in the peak temperature of the accretion flow. 

In summary, the parallels between HBC 340 and an EXor-type variable during outburst are inconsistent, with a lower 
mass accretion rate than anticipated being derived for HBC 340 and weaker H$\alpha$ emission after the protostar has 
brightened by nearly a magnitude. The outburst hypothesis, however, remains a valid possibility to explain the
gradual rise in brightness over the last two years.

\section{Summary}

We conclude that HBC 340 is primarily responsible for the brightness variations observed in the small, fan-shaped 
reflection nebula in NGC\,1333 that were reported by Hillenbrand et al. (2015). Our interpretation is that these 
brightness variations represent light echoes produced by varying incident radiation emanating from the near solar-mass 
protostar. The physical origin of variability in HBC 340, however, is difficult to isolate from the observations 
available due to the complexity of the system that includes at least two companions. Undoubtedly some fraction of 
variability arises from irregular accretion events, as evidenced by Balmer, Paschen, and Brackett emission 
as well as near infrared excess attributable to circumstellar dust. Additionally, correlated behavior on a 
several hundred day time scale appears in the light curve of the protostar that could indicate the presence of obscuring 
circumstellar material, possibly shepherded by HBC 340Ab, the near equal mass spectroscopic binary companion 
detected by high dispersion, optical spectroscopy. The low mass 
accretion rate estimated for HBC 340 could result from a cleared disk interior, swept up by the gravitational 
interaction of HBC 340Aa and HBC 340Ab. HST imaging has also revealed the presence of HBC 340B, a brown dwarf 
candidate lying $\sim$235 AU distant, with a mass between 20 and 50 times that of Jupiter, if associated with 
the protostar.

Over the course of two years, HBC 340 has brightened by more than two magnitudes at $R-$band, possibly the result 
of an EXor-like eruptive outburst. The estimated mass accretion rate of HBC 340, log $\dot{M}$$\sim$9.0, is not 
consistent with an EXor or FU Ori-type event, however. Such an increase in brightness may be characteristic of this 
stage of evolution as a Class I protostar transitions to a classical T Tauri star, clear of its natal envelope.
Alternatively, the rise in brightness could be attributable to the transit of an extended feature within its 
circumstellar disk, similar in nature to the long duration eclipse witnessed for LRLL 35 in IC\,348. This hypothesis, 
however, is not supported by the limited archival photometry of the source available from the literature, but it 
cannot be refuted either. It is also conceivable that extincting material along the line of sight to HBC 340 has 
been gradually removed over the years, perhaps by a protostellar outflow or jet. Continued photometric and spectroscopic 
monitoring of HBC 340 and its variable reflection nebula are warranted to further constrain the physical mechanisms 
responsible for their variability. Particularly valuable would be multi-band optical or near infrared photometry
that could constrain the color changes and test the extinction hypothesis. High angular resolution, polarimetric 
observations of the nebula would also prove insightful.

\acknowledgments

We have made use of the Digitized Sky Surveys, which were produced at the Space Telescope Science Institute under
U.S. Government grant NAG W-2166, the SIMBAD database operated at CDS, Strasbourg, France, and the 2MASS, a joint
project of the University of Massachusetts and the Infrared Processing and Analysis Center (IPAC)/California Institute
of Technology, funded by NASA and the National Science Foundation. This publication makes use of data products from
the Wide-field Infrared Survey Explorer, which is a joint project of the University of California, Los Angeles, and
the Jet Propulsion Laboratory / California Institute of Technology, funded by the National Aeronautics and Space
Administration. Some of the data presented herein were obtained at the W.M. Keck Observatory, which is operated as 
a scientific partnership among the California Institute of Technology, the University of California and the National 
Aeronautics and Space Administration. The Observatory was made possible by the generous financial support of the 
W.M. Keck Foundation. The Pan-STARRS1 Surveys (PS1) and the PS1 public science archive have been made possible through 
contributions by the Institute for Astronomy, the University of Hawaii, the Pan-STARRS Project Office, the Max-Planck 
Society and its participating institutes, the Max Planck Institute for Astronomy, Heidelberg and the Max Planck Institute 
for Extraterrestrial Physics, Garching, The Johns Hopkins University, Durham University, the University of Edinburgh, 
the Queen's University Belfast, the Harvard-Smithsonian Center for Astrophysics, the Las Cumbres Observatory Global 
Telescope Network Incorporated, the National Central University of Taiwan, the Space Telescope Science Institute, 
the National Aeronautics and Space Administration under Grant No. NNX08AR22G issued through the Planetary Science 
Division of the NASA Science Mission Directorate, the National Science Foundation Grant No. AST-1238877, the University 
of Maryland, Eotvos Lorand University (ELTE), the Los Alamos National Laboratory, and the Gordon and Betty Moore Foundation.
The authors gratefully acknowledge the efforts and dedication of the W. M. Keck Observatory staff and the NASA Infrared 
Telescope Facility staff for their support in making the observations presented herein. SED also wishes to thank Ethan 
Dahm for assisting with the HIRES observations made on 30 October 2015. Finally the authors thank Bo Reipurth who served 
as the referee for this paper. His insightful comments and recommendations greatly improved the manuscript.

\vspace{6cm}

\begin{deluxetable}{ccc}
\tabletypesize{\small}
\tablewidth{0pt}
\tablenum{1}
\tablecaption{Observed and Derived Properties of HBC 340 and HBC 341}
\tablehead{
\colhead{Property}   &    \colhead{HBC 340} & \colhead{HBC 341}\\}
\startdata
$V$ (mag)\tablenotemark{a}         &  17.78$\pm$0.003   &    19.66$\pm$0.02 \\
$V-R_{C}$ (mag)\tablenotemark{a}   &   1.33$\pm$0.003   &     1.79$\pm$0.03 \\
$V-I_{C}$ (mag)\tablenotemark{a}   &   2.71$\pm$0.003   &     4.38$\pm$0.02 \\
$A_{V}$ (mag)     &   8.0$\pm$1.0    &     3.5$\pm$1.0  \\
$M_{V}$ (mag)     &   2.92$\pm$1.0   &     9.30$\pm$1.0 \\
Spectral Type     &    K7$\pm$1    &      M5$\pm$1  \\
T$_{eff}$ (K)\tablenotemark{b}     &   3970$\pm$100   &     2880$\pm$100 \\
Mass (M$_{\odot}$)\tablenotemark{c} &  1.0$\pm$0.25    &     0.26$\pm0.10$ \\
Age (Myr)\tablenotemark{c}         &   0.1+1.0     &     2.0$\pm$1.0   \\
\enddata
\tablenotetext{a}{Photometry is from the UH 2.2 m observations in October 1999.}
\tablenotetext{b}{From the pre-main sequence temperature scale of Pecaut \& Mamajek (2013).}
\tablenotetext{c}{Derived from the Siess et al. (2000) pre-main sequence models.}
\end{deluxetable}

\begin{deluxetable}{cccc}
\tabletypesize{\small}
\tablewidth{0pt}
\tablenum{2}
\tablecaption{Measured Equivalent Widths of Emission Features in HBC 340 and HBC 341}
\tablehead{
\colhead{Transition}   &    \colhead{HBC 340 (OCT 2015)} & \colhead{HBC 341 (OCT 2015)} & \colhead{HBC 340 (JAN 2017)}\\
            &    W(\AA)\tablenotemark{a}       &    W(\AA)      &   W(\AA)    \\}
\startdata
Keck HIRES           &                 &             &              \\
\hline
H$\beta$             &    $-$1.45      &   $-$25.61  &        ecr\tablenotemark{b}   \\
Fe II $\lambda$5158  &    $-$0.62      &    NP\tablenotemark{b}       &        NP   \\
O I $\lambda$5577    &    $-$0.17      &   $-$0.47   &        NP   \\
O I $\lambda$6300    &    $-$5.25      &   $-$3.52   &      $-$2.4  \\
O I $\lambda$6363    &    $-$1.31      &   $-$0.80   &      $-$0.5  \\
H$\alpha$            &    $-$10.93     &   $-$31.35  &      $-$5.9  \\
N II $\lambda$6583   &    $-$0.83      &   $-$1.34   &      $-$0.6  \\
S II $\lambda$6717   &    $-$0.43      &   $-$1.48   &      $-$0.2  \\
S II $\lambda$6731   &    $-$1.42      &   $-$1.52   &      $-$0.7  \\
Ca II $\lambda$8542  &    ecr\tablenotemark{b} &  $-$0.41 & ecr\tablenotemark{b} \\
\hline
IRTF SpeX            &                 &             &              \\
\hline
He I $\lambda$10830  &    $-$5.75      &   abs\tablenotemark{b}    &       NA\tablenotemark{b}   \\
Fe II $\lambda$12570 &    $-$0.90      &    NP\tablenotemark{b}    &       NA\tablenotemark{b}   \\
Pa$\beta$            &    $-$0.92      &    NP\tablenotemark{b}    &       NA\tablenotemark{b}   \\
Br$\gamma$           &    $-$0.53      &    NP\tablenotemark{b}    &       NA\tablenotemark{b}   \\
\enddata
\tablenotetext{a}{Negative values imply emission.}
\tablenotetext{b}{ecr-emission core reversal; NP-feature not present; abs-absorption; NA-observations not available.}
\end{deluxetable}

\begin{deluxetable}{cccc}
\tabletypesize{\small}
\tablewidth{0pt}
\tablenum{3}
\tablecaption{Measured Radial Velocities of HBC 340}
\tablehead{
\colhead{UT Date}   &    \colhead{Systemic} & \colhead{Negative Component (A)} & \colhead{Positive Component (B)}\\
            &    (km s$^{-1}$)       &    (km s$^{-1}$)      &   (km s$^{-1}$)    \\}
\startdata
27 October 2015    &    17.32$\pm$3.85      &  0.68$\pm$4.47  & 35.54$\pm$2.03  \\
30 October 2015    &    18.17$\pm$5.35      &  5.01$\pm$0.51  & 31.28$\pm$1.72  \\
12 January 2017    &    14.76$\pm$2.93      &  $-$1.66$\pm$2.87 & 35.83$\pm$2.20 \\
\enddata
\end{deluxetable}

\end{document}